# What is measured in a photoluminescence experiment on Quantum dots embedded in a large Purcell factor microcavity?

B. Gayral and J.M. Gérard

CEA-CNRS group "Nanophysique et Semiconducteurs", CEA-Grenoble, INAC/SP2M, 17 rue des Martyrs, 38054 Grenoble, France

**Abstract:**

It is usually assumed that when performing a photoluminescence experiment on a microcavity containing an inhomogeneously broadened quantum dots ensemble, the cavity mode appears as a positive peak with a linewidth that reflects the mode quality factor Q. We show in this article that this conclusion is in general not true, and that the measured mode linewidth depends strongly on the excitation power for microcavities having large Purcell factors. We analyze theoretically this effect in the case of the micropillar cavity and we show that the same microcavity can give rise to a large variety of photoluminescence spectral signatures depending on the excitation power and collection set-up. We finally give guidelines to measure the real cavity quality factor by photoluminescence.

## I Introduction

For more than ten years, self-assembled quantum dots (QDs) have been recognized as an emitter of choice to study microcavity effects in semiconductor systems. One reason firstly shown in micropillar cavities [1] is that their broad homogeneous emission linewidth allows to probe microcavity modes in a simple photoluminescence (PL) experiment over a wide spectral range (typically 100 meV for InAs/GaAs QDs). Unlike quantum wells, the absorption induced by the QD array is in general negligible compared to the optical losses of the empty cavity. The measurement of the mode energies and quality factors using QDs as an internal broad-band light source was successfully performed for micropillars [1], microdisks [2] and various photonic crystal structures [3 - 6]. The demonstration of a large Purcell effect for QDs in micropillars [7] opened later the way to numerous solid-state cavity quantum electrodynamics experiments (for a review see [8]), including the vacuum Rabi flopping of a single QD [9 - 11], with promising prospects for low-threshold lasers [12], single photon sources [13] and possibly quantum information processing [14]. In this article, we come back to a basic questions that oddly enough was not raised yet in the literature : what is exactly measured when a PL experiment is performed on a microcavity containing quantum dots? Since the first publication on the topic [1], it has been assumed that the cavity modes appear as positive peaks in the spectrum, and that the linewidth of the PL peaks faithfully reflects their quality factor Q. Obviously, the QD-cavity system should at least satisfy two conditions for that. First of all, the number of QDs should be large enough to ensure a good "smoothness" of the spectrum of the internal light source. This condition simply writes $\Delta E_{\text{hom}} \times N \gg \Delta E_{inh}$, where N is the total number of QDs within the mode area, $\Delta E_{\text{hom}}$ the homogeneous linewidth of the QD transitions and $\Delta E_{\text{inh}}$ the inhomogeneous linewidth of the QD distribution. Secondly, the additional absorption losses introduced by the QD array should be much smaller than the empty cavity losses. We show in this paper that even when both conditions are fulfilled, the PL peaks *do not* reflect properly the Q's of the cavity modes in general, when the QDs experience a strong Purcell effect. More precisely, we show that the PL peak can be broadened by a factor as large as $(F_p+1)^{-1/2}$ compared to the cavity mode, for a weak excitation of the QD array ($F_p$ is the Purcell factor of the cavity). We also show that in the Purcell-enhancement regime, cavity modes can appear as dips as well as peaks in the PL spectra, depending on the experimental configuration that is used for photon collection. These effects are analyzed in more detail on the basis of a simple model accounting for the spatial dot distribution inside the cavity (the micropillar is taken as an example, although the results apply to all cavities) and the level structure of the QDs (excitonic and biexcitonic transitions). This analysis allows to select several experimental protocols for a reliable measurement of the cavity Q in a PL experiment using QDs as internal light source.

The paper is organized as follows. In part II, we present the theoretical basis of the paper, and demonstrate that the mode PL peak linewidth depends strongly on excitation power and reflects the actual cavity Q only in the high excitation power limit. In part III, we analyze the various spectra that can be obtained in PL on a microcavity containing QDs as a function of the collection set-up. We discuss in part IV the limits of our model and the various strategies that can be implemented to measure correctly in PL the quality factor of a large Purcell factor microcavity.

## II PL linewidth of cavity modes in the Purcell-enhancement regime

The geometry that we consider here is a GaAs/AlAs micropillar with circular section containing a GaAs λ-cavity with an InAs QD array at its center. We assume that the pillar diameter is small enough for resonant modes to be spectrally well separated, and focus our attention on the coupling of QDs to the fundamental mode $HE_{1,1}$, which is doubly polarization degenerate. Note that the results derived hereafter can be generally applied to any microcavity geometry. For pedagogical purposes, we first consider a hypothetical situation where in the plane perpendicular to the micropillar axis, all the QDs are located at the center of the pillar i.e. at the spatial maximum of the mode electromagnetic field. This is not a realistic situation, but in that case the calculations are analytical and contain all of the relevant physics. The mode has a quality factor Q and an effective volume V (see [13] for a definition of the effective volume) and resonates at energy $E_0$. We suppose in the entire article that the QD density is sufficiently low and the quality factor sufficiently small so that reabsorption of cavity photons by the QDs can be neglected. The Purcell factor [15] is given by $F_p = \frac{3}{4\pi^2} \frac{Q(\lambda/n)^3}{V}$, where n is the refractive index of GaAs. It is assumed that the QDs have an in-plane dipole, as observed experimentally for their fundamental interband optical transition [16]. Finally, we assume that the experiment is performed at low temperature under weak excitation conditions, so that the emission spectrum of a single QD consists in a single narrow excitonic emission line.

The radiative emission rate for a quantum dot emitting at energy E and located at the center of the pillar is then given by:

$$\Gamma(E) = \Gamma_0 (F_p \frac{E_0^2}{4Q^2(E-E_0)^2 + E_0^2} + \gamma) = \Gamma_0(F_p L(E) + \gamma) \quad (1)$$

where $\Gamma_0$ is the radiative emission rate of the InAs QDs in an homogenous GaAs matrix, ($\gamma.\Gamma_0$) is the emission rate in the leaky modes of the pillar, and L(E) is the lorentzian spectral distribution of the fundamental cavity mode. In general, $\gamma$ is slightly less than unity when the QD emits within the Bragg reflector stop band ($\gamma \sim 0.8$ [7]) due to the fact that the spectral density of leaky modes seen by a QD in a micropillar is less than the density of modes in a homogeneous matrix. This effect is however quite small and negligible in the case of large $F_p$ microcavities that we study in this article, so that we shall take here $\gamma=1$ for all numerical examples. Note that for photonic crystal cavities, $\gamma$ can be significantly smaller than 1 [17]. Equation (1) gives the total QD spontaneous emission rate, taking into account the fact that the micropillar modes are polarization degenerate, the partial emission rates being $\Gamma_0 F_p L(E)$ in the cavity mode and $\gamma \Gamma_0$ in the leaky modes. The fraction of the QD spontaneous emission that is emitted in the mode, $\beta(E)$, can be written as $\beta(E) = \frac{\Gamma_0 F_p L(E)}{\Gamma(E)} = \frac{F_p L(E)}{F_p L(E) + \gamma}$.

We now estimate the number of photons that are actually detected by the detector in the PL experiment. Let us call A and B the collection plus detection efficiencies for photons emitted in the cavity mode and in the leaky modes respectively. Parameters A and B depend on the detection set-up. If a given QD emits $I_{QD}(E)$ photons per second, then the detected signal coming from this particular QD is :

$$I(E) = A I_{QD}(E) \frac{\Gamma_0 F_p L(E)}{\Gamma(E)} + B I_{QD}(E) \frac{\gamma \Gamma_0}{\Gamma(E)} = A I_A(E) + B I_B(E) \quad (2)$$

The first term represents the emission in the mode, which appears as a positive peak in the PL spectrum, while the second term represents the emission into the leaky modes.

In general when collecting photons stemming from a micropillar cavity along the pillar axis, A is much larger than B (this is also true for other microcavities as long as the collection set-up is well adapted to collect the far-field emission stemming from the mode) so that in a PL experiment the mode signal (first term) appears as an intense positive peak against a weak background (second term). Let us then focus in this part on the first term related to the mode emission. By developing L(E), $I_A(E)$ can be rewritten as :

$$I_A(E) = AI_{QD}(E) \frac{F_p}{F_p + \gamma} \frac{E_0^2}{4 \frac{\gamma Q^2}{F_p + \gamma}(E - E_0)^2 + E_0^2} \quad (3)$$

If now one assumes that all QDs are excited at the same pump rate, that the quantum emission yield is unity for all QDs and that all QDs are well below saturation (i.e. the probability to have a biexciton in the QD is vanishingly small), then $I_{QD}(E)$ is independent of E and equal to the pumping rate per QD $P_{exc}$. All these assumptions hold when exciting the QDs non-resonantly at low temperature and low enough power. In these conditions, the PL spectrum reflects the spectral dependence of the spontaneous emission coupling factor in the mode β(E). One then sees from equation (3) that the collected signal from the mode has a lorentzian spectral shape, however with a linewidth $\frac{E_0}{Q}\sqrt{\frac{F_p + \gamma}{\gamma}}$. This means that as soon as the Purcell factor becomes of the order of γ, the measured linewidth at low power does not reflect the quality factor of the mode but an apparent Q value which is smaller by a factor $\sqrt{\frac{F_p + \gamma}{\gamma}}$. We also point out that the same analysis should be applied to determine the spectral dependence of the PL intensity collected from a single QD being tuned in and out of resonance with a cavity mode [18, 19].

The physical explanation behind this phenomenon is actually quite intuitive. In the limit of vanishingly small Purcell factor, β(E) is proportional to L(E) so that the spectral shape of the collected PL reflects the spectral density of states of the confined mode. On the opposite in the limit of a strong Purcell factor, β(E) becomes close to one even for QDs that are quite off-resonance with the mode peak. Thus all QDs that couple reasonably well to the mode emit practically all of their photons in the cavity mode, which broadens the spectrum of the PL signal stemming from the mode.

As discussed earlier, the situation depicted here is not realistic. In the experiments that are actually performed, the QDs are spatially distributed all over the section of the micropillar so that they all couple differently to the mode. In order to account for this spatial QD distribution, one can rewrite equation (2) as:

$$I(E,\vec{r}) = AI_{QD}(E,\vec{r}) \frac{F_p L(E)\left|\vec{E}_{xy}(\vec{r})\right|^2}{F_p L(E)\left|\vec{E}_{xy}(\vec{r})\right|^2 + \gamma} + BI_{QD}(E,\vec{r}) \frac{\gamma}{F_p L(E)\left|\vec{E}_{xy}(\vec{r})\right|^2 + \gamma} = AI_A(E,\vec{r}) + BI_B(E,\vec{r}) \quad (4)$$

where $\vec{E}_{xy}(\vec{r})$ is the in-plane electric field at the location of the QD. In order to obtain the collected spectrum, one has to sum over all QDs that spatially couple to the mode:

$$I(E) = \iint_{\substack{Mode \\ area}} \left( AI_{QD}(E,\vec{r}) \frac{F_p L(E)\left|\vec{E}_{xy}(\vec{r})\right|^2}{F_p L(E)\left|\vec{E}_{xy}(\vec{r})\right|^2 + \gamma} + BI_{QD}(E,\vec{r}) \frac{\gamma}{F_p L(E)\left|\vec{E}_{xy}(\vec{r})\right|^2 + \gamma} \right) d\vec{r} = AI_A(E) + BI_B(E) \quad (5)$$

In that case, the linewidth of the PL peak is spectrally broader than the cavity mode by a factor $\sqrt{\frac{F_p^{eff} + \gamma}{\gamma}}$, where the effective Purcell factor $F_p^{eff}$ is obtained by averaging the magnitude of the Purcell effect over all possible positions of the QDs within the mode area (it is assumed that the QD density is spatially uniform). The ratio between the effective Purcell factor and the microcavity Purcell factor is of the order of 3 to 4 and depends on the micropillar diameter and quality factor [7]. For instance for a micropillar with Q=2300 and diameter 1 µm, the Purcell factor is 28 and the effective Purcell factor (as defined in this article) is 8.6. This means that under low excitation conditions, a PL linewidth corresponding to a quality factor of around 700 would be measured on such a micropillar.

From these results two questions arise. i) What happens when the excitation power is raised so that the excitonic and then biexcitonic transitions can be saturated? ii) Is the real Q measured in the many papers published so far with microcavities containing QDs? These questions are actually linked and we shall now study the PL behavior of a micropillar containing QDs when the excitation power is raised. The behavior of QDs as a function of excitation power is actually quite complicated [20]. Beyond the well known exciton and biexciton transitions, many particle complexes form at high power. This effect is particularly important in the case of high $F_p$ cavity : at the power needed to saturate the excitonic transition of a QD on resonance with the mode, an off-resonance QD is well beyond saturation for its s shell levels. In the present case, we restrict ourselves to a simple model taking into account only the excitonic and biexcitonic transitions. In other words, we discard the effect of p-shell (and higher lying shells) carriers on the s-shell excitonic and biexcitonic transitions. While this model will not reproduce exactly the PL spectra as a function of pumping power over orders of magnitudes, it gives the main results, i.e. the difference between low power excitation where all QDs emit the same number of photons (pinned by the homogeneous excitation level) and high power excitation (i.e. when the s shell QD levels are fully occupied), for which the number of photons emitted by a QD on s shell transitions is governed by its Purcell-enhanced spontaneous emission rate. We also assume that within the excitation power range that we consider, the homogeneous linewidth of the QD transitions remains small enough (i.e. much smaller than the cavity linewidth) so that the Purcell effect is not degraded [21]. This can be for instance obtained by exciting resonantly the bottom of the wetting layer [20]. For the model, we assume that the biexciton binding energy is 3 meV, with a gaussian distribution with FWHM of 0.6 meV [22]. The rate equations we use for a given QD are the following:

$$\frac{dg}{dt} = \gamma_x X - Pg$$
$$\frac{dX}{dt} = \gamma_{x_2} X_2 - PX - \gamma_x X + Pg \qquad (6)$$
$$\frac{dX_2}{dt} = -\gamma_{x_2} X_2 + PX$$

where g is the probability to be in the ground state (empty dot), X the probability to have a single exciton and $X_2$ the probability to have a biexciton. $\gamma_X$ and $\gamma_{X_2}$ are the spontaneous emission rates of the exciton and biexciton respectively (taking into account the Purcell enhancement). P is the pumping rate per QD, assumed to be the same for all QDs. Solving for these equations in the steady-state regime yields:

$$I_x = \frac{P}{1 + \frac{P}{\gamma_x} + \frac{P^2}{\gamma_x \gamma_{x_2}}}$$
$$I_{x_2} = \frac{P^2}{\gamma_x \left(1 + \frac{P}{\gamma_x} + \frac{P^2}{\gamma_x \gamma_{x_2}}\right)} \qquad (7)$$

where $I_X$ and $I_{X_2}$ are the number of photons emitted per time unit by the excitonic and biexcitonic transitions respectively. One can then compute the number of photons emitted in the microcavity mode $I_A$ and in the leaky modes $I_B$, as a function of pump power. It is assumed here that for a QD in an homogeneous GaAs matrix, $\gamma_X = \Gamma_0 = \gamma_{X_2}/2$. Fig. 1 displays $I_A$ (emission in the mode) for a micropillar with diameter 1 µm and Q = 15000 (Fp= 189). The measured Q varies from 2200 at low power (P=0.01 $\Gamma_0$) to 13700 at high power (P=1000 $\Gamma_0$). The difference between the low power spectrum and the high power spectrum is thus very large.

In figure 2, we display the "measured Q" (i.e. PL peak energy divided by the PL peak linewidth) as a function of pumping power for two different micropillars (including the one discussed in figure 1). Worth noticing, the measured Q value increases and saturates at the real Q value in the limit of strong excitation. This behavior can be understood as follows : in the strong excitation limit, the number of photons that can be emitted by a biexcitonic transition is proportional to its total spontaneous emission rate. From equation (2) one sees that the total emission in the mode is proportional to the Purcell enhancement and thus reflects L(E).

Let us note that measuring such linewidth dependencies as a function of power is an elegant way of measuring the average magnitude of the Purcell effect in a simple continuous wave (cw) PL experiment. This approach is complementary to the kind of experiments that have already been performed by looking at the different saturation behaviors for off- and on-resonance QDs under cw excitation to probe the Purcell effect [23 - 25].

We have thus seen in this part that in a microcavity containing QDs in the Purcell-enhancement regime, the measured quality factor for the cavity mode in a PL experiment drastically depends on the excitation power. It is only at high excitation power that the actual quality factor is measured.

**III The role of the collection geometry**

So far, we have assumed that the mode emission is collected preferentially. However one can wonder what happens when the leaky modes are collected preferentially. Figure 3 represents the leaky modes intensity $I_B$ for the same micropillar studied in part II in the same power range. The cavity mode appears here as a negative peak, which makes sense : as the QDs coupled to the cavity have a high β factor, they emit most of their photons into the cavity mode and not into the leaky modes. In other words at low excitation power, the PL spectrum reflects (1-β(E)). This spectral feature which is a simple demonstration of the Purcell effect in a cw PL experiment has oddly never been reported. This requires developing a set-up which collects as few cavity mode photons as possible, which is doable for a cavity mode which emits very directionally (for instance a microdisk cavity emitting only in a narrow cone around the disk plane [26]). To pursue the various collection possibilities, which affect drastically the measured spectra, we envision in figure 4 a hypothetical experiment where all photons are collected in all directions (this corresponds to A=B=1). In the absence of Purcell effect, the spectra should be completely flat. This is also what is seen in the limit of low power : in that case all QDs emit the same number of photons (pinned by the pumping power), and as they are all collected and detected with the same efficiency in that configuration, the spectrum does not show any structure. As soon as some transitions start to saturate, while others do not due to the Purcell effect, the spectrum becomes structured. At high power where all transitions are saturated, the cavity mode appears as a positive peak as transitions (in that case biexcitonic transitions) on resonance with the mode can emit more photons at saturation.

In general, a realistic experiment would have a collection situation where A and B are not equal. When studying microcavities, the experiment favors a situation where A>B to collect efficiently the cavity mode photons. In that case, the spectra are very much like the one in figure 1, as the mode emission dominates over the leaky modes background (note that even in the case where A=B, one recovers a situation where the mode dominates the spectrum for P above 10 $\Gamma_0$). A different situation is the one where the leaky modes are collected preferentially, while the mode photons are still collected (which is a more realistic situation than figure 3 for which no cavity photons are collected). Figure 5 shows the power dependent spectra for B=10A. In that case at low power the spectrum is dominated by the cavity mode "dip" due to the less favorable collection of on-resonance photons, while at high power it is dominated by the collected mode emission despite the low collection factor A.

We therefore conclude that the PL spectrum of a QD array in a microcavity in the Purcell-enhancement regime depends both on the pumping power and on the collection geometry, especially if it disfavors the collection of cavity mode photons. Studying the PL spectrum as a function of excitation power and collection set-up is thus a powerful way of fully characterizing the microcavity by measuring both the quality factor and demonstrating the Purcell effect in a simple cw experiment.

## IV Discussion

### A- Validity and limits of the model

We shall now come back to the validity of our model over the entire power range that we simulate. The assumptions that we have made are as follows : a) we have taken a simple model of QD state filling taking into account only the excitonic and biexcitonic transitions and b) we have assumed that the homogeneous linewidths of the QD transitions remain small compared to the cavity mode linewidth. We also have assumed that absorption of the cavity mode photons by the QDs is negligible, and we will not further discuss this as it would be way beyond the scope of this article to cope with the issue of an absorbing emitter. Concerning assumption a), the limitation of our model is the difficulty of modeling precisely the emission dynamics of the QDs including all multiparticle states over such a large power range. Our calculations are of course correct in the low power limit where no transition saturates. The other interesting case is the high power limit. In that case, as shown for instance in reference [20], when exciting strongly a collection of QDs (orders of magnitudes higher than the exciton saturation), the s-shell emission saturates with an unchanged spectrum compared to the low power limit. At very high power, the s-shell emission is made of the transitions of biexcitons dressed by various p-states filling configuration. It can be assumed that these various transitions all have the same decay time as the p-state dressing will shift transition energies but is not likely to significantly affect oscillator strengths of the s-shell transitions. Coming back to an ensemble of QDs excited at very high power and coupled to a cavity mode, various QD dressed biexcitonic transitions will feed the mode, each of them shifting spectrally with the variation of p-state occupation. Averaged over all the QDs, the mode is fed by a constant number of saturated transitions that all have similar radiative recombination rates, and we thus are in the same situation as in our simple model. We thus can say that in figure 2, the low power part (below ~$\Gamma_0$) and the high power part (above ~100 $\Gamma_0$ for both cases considered here, although the limit for this validity zone of course depends on the cavity Purcell factor) are correct despite the limitations of our model. There can however be discrepancies between actual experiments and our simple model in the intermediate power zone. In particular in this intermediate range of excitation power, what can happen is that QDs for which neither the excitonic nor the biexcitonic lines are in resonance with the mode have p-states dressed biexcitonic transitions that are in resonance with the mode. This effect will certainly affect the mode intensity as more transitions will feed the mode compared to what is forecasted by our model. This will however marginally affect the mode spectral shape as these dressed biexcitonic transitions behave like regular biexcitonic transitions in terms of lifetime and saturation behavior.

Assumption b) can be valid when exciting resonantly the bottom of the wetting layer [20]. However in most cases when exciting higher in energy, the QD transitions broaden considerably at high excitation power [27] so that their homogeneous linewidth can become larger than the cavity mode linewidth. This then degrades the Purcell enhancement [21] so that the true cavity mode linewidth is measured in PL as in a low Purcell factor cavity. When this type of excitation power induced broadening occurs, the shape of the curves in figure 2 will still be the same, only the transition to the high power regime where the real cavity Q is measured will occur for smaller excitation powers due to the degradation of the Purcell enhancement that occurs as the power is raised.

**B- Guidelines to measure the real cavity Q**

If we now come back to the two initial questions asked at the beginning of this article, question i) can now be answered : when saturating all transitions, one recovers a PL spectrum giving the real cavity Q. Question ii) is more difficult to answer, mainly because there are few experiments in which the pumping power is mentioned when measuring a microcavity mode PL spectrum. We assume that for most articles that measure a cavity quality factor by performing a PL experiment on a microcavity containing QDs, the experimentalist favored high power excitation in order to increase the collected signal as well as the mode to background (i.e. leaky modes) ratio, so that it is expected that most articles in the literature did measure the correct quality factor.

More generally, one might wonder how to measure precisely a microcavity quality factor, without having to worry about the precise excitation power that is used. As discussed in this article, the two phenomena that are complex to deal with are firstly the large Purcell factor that leads to spectrally dependent saturation phenomena for the QDs and secondly the power dependent density of states of an inhomogeneously broadened ensemble of QDs. Both phenomena can be suppressed by raising the sample temperature : in that case, the QD homogeneous broadening can be as large as 5 meV for InAs QDs at room temperature [28]. In such a regime, for which the emitter linewidth is larger than the mode linewidth, the magnitude of the Purcell effect must be calculated using the quality factor of the emitter ($Q_{em}$ =250 for a 5 meV emission linewidth around 1.3 eV) instead of the cavity Q [21]. This leads to a decrease of the Purcell effect [21] which will in most cases (except for very low volume cavities) reduce considerably the effects described above. For such experimental conditions, the PL peak linewidth will be only weakly power dependent and will thus give a reliable estimate of the cavity quality factor.

**V Conclusion**

As a conclusion, we have shown that the PL spectra measured on an ensemble of QDs embedded in a large Purcell factor microcavity depend strongly on the excitation conditions and on the collection set-up. It is thus very important to have a good understanding of the experimental conditions when measuring a quality factor in PL. The correct quality factor is in particular measured in the limit of strong excitation power when all transitions are saturated. By analyzing the measured quality factor variations when changing the excitation power, it is possible to demonstrate the Purcell effect in a simple cw PL experiment. Moreover, proof of the Purcell effect can also be obtained by observing a dip in the PL spectrum when collecting mostly photons emitted in the leaky modes. Finally we point out that a reliable protocol to measure the Q factor by PL consists in decreasing the Purcell enhancement by increasing the homogeneous broadening of the QD transitions. This can be obtained by raising the sample temperature and/or by exciting at high enough power.

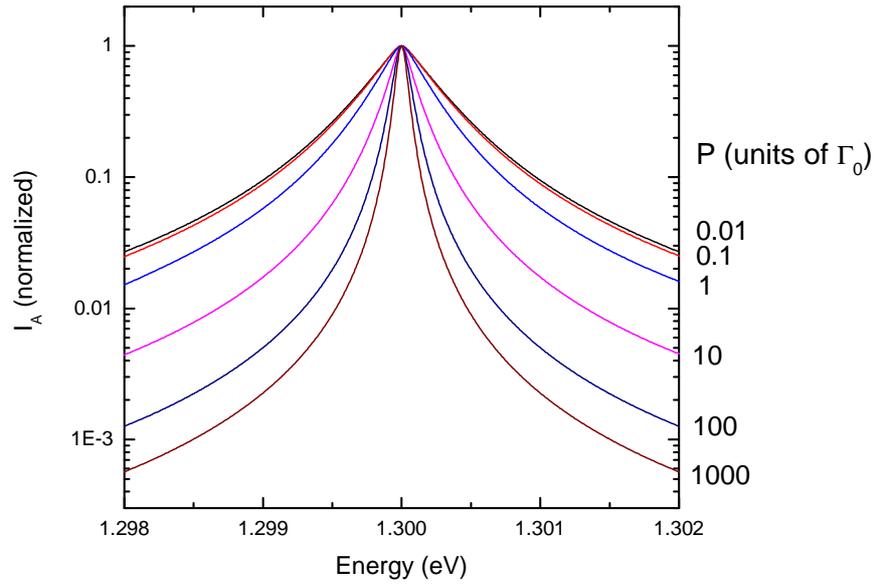

Figure 1 : Normalized spectra for the "mode" photons for a micropillar with diameter 1 μm and Q factor 15000 as a function of pumping power. The decrease of the measured linewidth when the pump power is increased appears clearly.

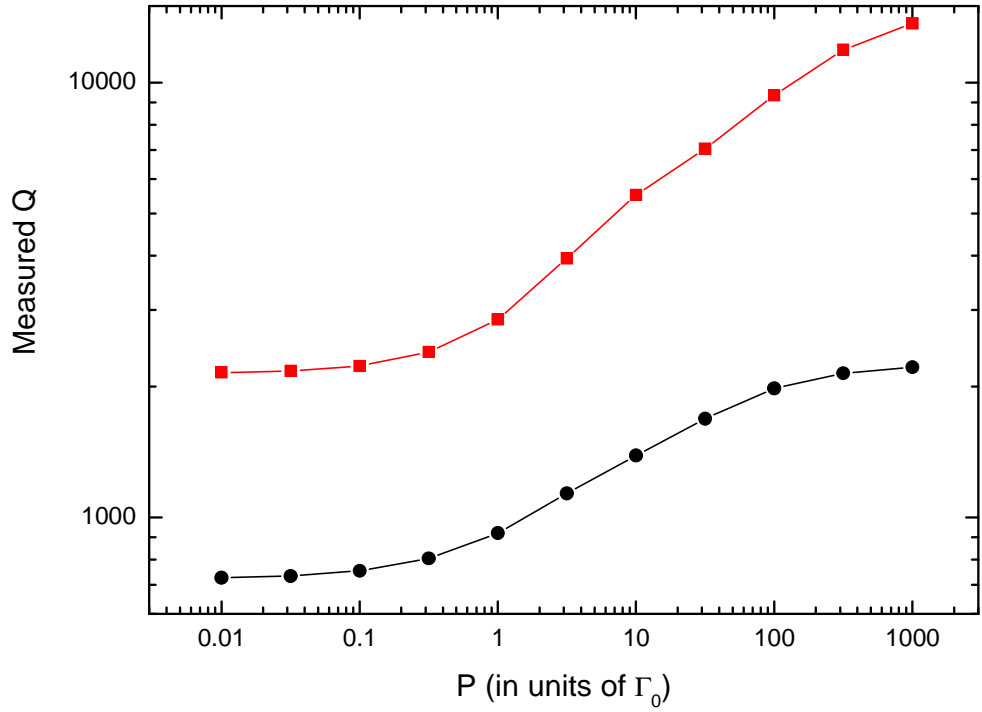

Figure 2 : Power dependence of the measured Q for micropillars with a diameter of 1 µm, and quality factors of 2300 (black circles) and 15000 (red squares).

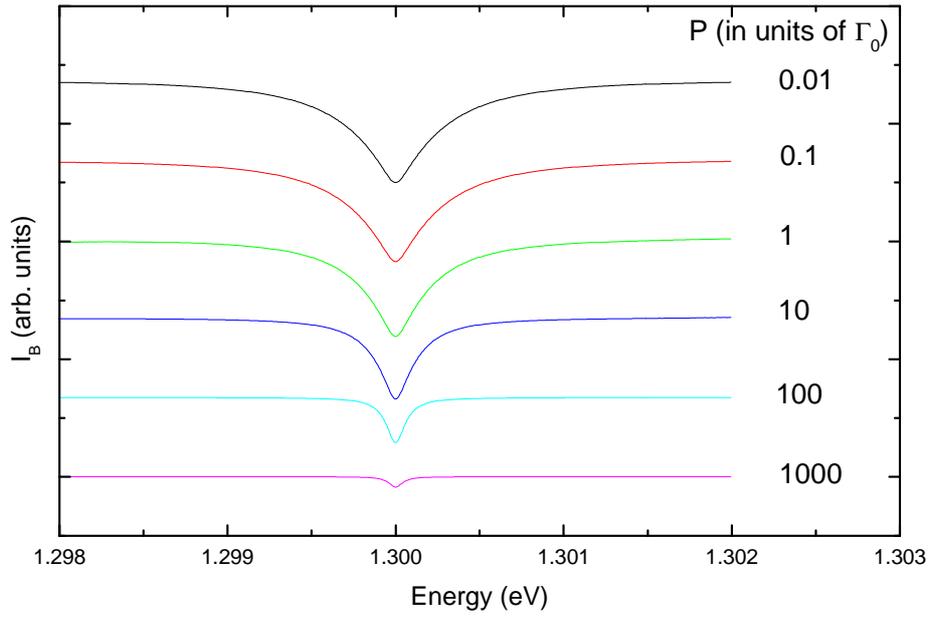

Figure 3 : Leaky modes intensity as a function of pumping power for a pillar of diameter 1 µm and Q factor 15000. The spectra are vertically offset for clarity.

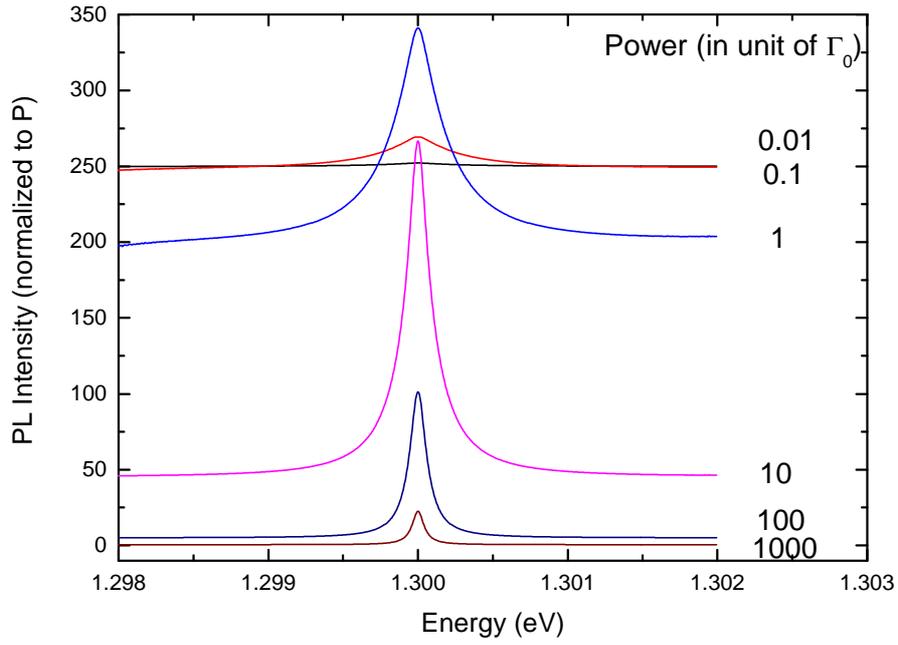

Figure 4 : Collected spectrum for a micropillar with a diameter of 1 µm and a Q of 15000 for a detection of all emitted photons (A=B=1). The spectrum for P=0.01 is the flattest one.

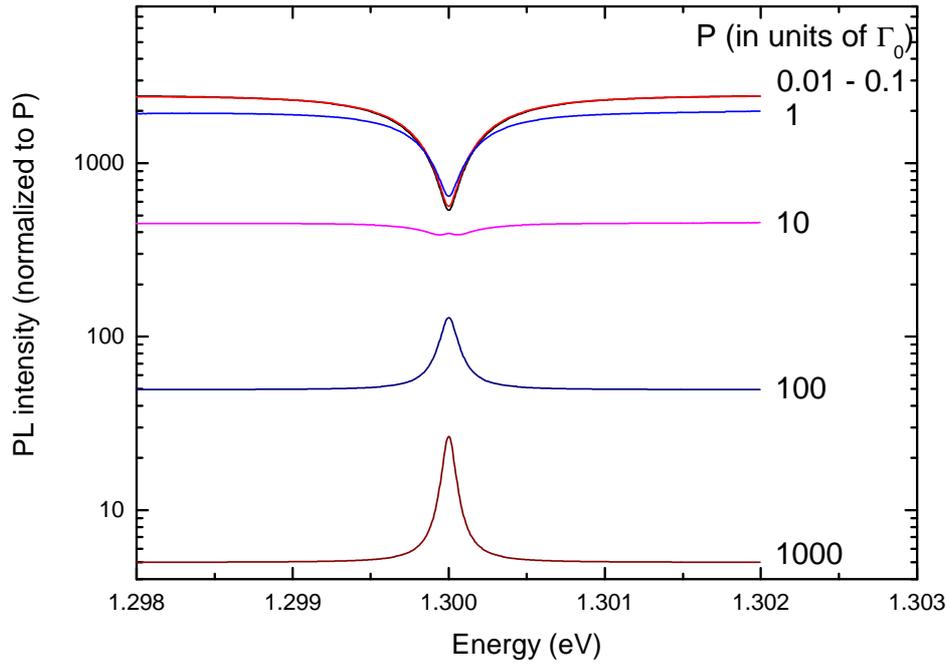

Figure 5 : Collected spectra for a micropillar with a diameter of 1 µm and a Q of 15000 as a function of power (normalized to P) for collection factors B=10A.